# Hyperbolic Growth of the World Population in the Past 12,000 Years

Ron W Nielsen[1]

Environmental Futures Research Institute, Gold Coast Campus, Griffith University, Qld, 4222, Australia

Data describing the growth of the world population in the past 12,000 years are analysed. It is shown that, if unchecked, population does not increase exponentially but hyperbolically. This analysis reveals three approximately-determined episodes of hyperbolic growth: 10,000-500 BC, AD 500-1200 and AD 1400-1950, representing a total of about 89% of the past 12,000 years. It also reveals three demographic transitions: 500 BC-AD 500, AD 1200-1400 and AD 1950-present, representing the remaining 11% of the past 12,000 years. The first two transitions were between sustained hyperbolic trajectories. The current transition is to an unknown trajectory. There was never any form of dramatic transition from stagnation to growth, described often as a takeoff, because there was no stagnation in the growth of the world population. Correct understanding of the historical growth of human population is essential in the correct interpretation of the historical growth of income per capita.

**Introduction**

The study of the historical economic growth involves not only the study of the Gross Domestic Product (GDP) but also the study of the growth of the population, because as pointed out by Galor (Galor, 2005, 2011, 2016), it is important to understand the relationship between the these two process and particularly the relationship between the growth of the income per capita (GDP/cap) and the growth of the population. It is, perhaps, for this reason that the latest and the most extensive compilation of the historical GDP data, published by the world-renown economist, includes also the data describing the historical growth of human population (Maddison, 2001, 2010).

About 50 years ago, von Foerster, Mora and Amiot (1960) demonstrated that human population was increasing hyperbolically during the AD era. We now have far better and more extensive sets of data compiled not only by Maddison (2001, 2010) but also by Manning (2008) and by the US Census Bureau (2016). The last two compilations are based on virtually the same primary sources but they are complimentary.

Maddison's compilation is useful in studying the growth of the population not only global but also regional and national. However, his data are terminated in AD 1. Furthermore, they also contain significant gaps below AD 1500. The data compiled by Manning and by the US Census Bureau are significantly richer but they are limited only to the description of the world population. However, they extend down to 10,000 BC.

It is well outside the scope of the discussion presented here, but a preliminary examination of Maddison's data indicates that the economic growth and the growth of human population followed similar trajectories. Consequently, by using a rich set of data extending down to 10,000 BC we might gain a better insight not only into the historical growth of human population but also to its possible link with the economic growth.

---

[1] AKA Jan Nurzynski, r.nielsen@griffith.edu.au; ronwnielsen@gmail.com

**The data**

Procedures adopted in estimating historical populations are described by Durand (1977). The data for the AD era are of exceptionally good quality. Between AD 400 and 1850, independent estimates are within ±10% of their corresponding averaged values. The estimates after 1850 are within ±1.5%. The largest deviations of around ±30% are for the AD 1 data. The two estimates for AD 200 differ by ±15% from their average value. The BC data are less accurate and less consistent but when closely analysed they are also found to follow a certain, well-described trajectory.

**Analysis of population data**

In order to understand hyperbolic distributions it is useful to compare them with the more familiar exponential distributions. The differential equation describing exponential growth is given by the following simple equation:

$$\frac{1}{S(t)}\frac{dS(t)}{dt} = k, \tag{1}$$

where $S(t)$ is the size of a growing entity, in our case the size of the population, and $k$ is an arbitrary constant.

The left-hand side of this equation represents growth rate. For $k > 0$ the eqn (1) describes growth, while for $k < 0$ it describes decay.

The solution of the eqn (1) is

$$S(t) = ae^{kt}, \tag{2}$$

where $a$ is the constant related to the constant of integration.

The eqn (2) gives

$$\ln S(t) = \ln a + kt. \tag{3}$$

The logarithm of the size of the growing entity increases linearly with time. Exponential growth can be easily identified by plotting data using semilogarithmic scales of reference because in such presentation the data should follow an increasing straight line.

Data for the growth of the population during the BC and AD eras (Manning, 2008; US Census Bureau, 2015) are shown in Figure 1. They are compared with the best exponential fit to the data. The world population was not increasing exponentially.

Let us now examine the hyperbolic growth. This type of growth is described by the following differential equation:

$$\frac{1}{S(t)}\frac{dS(t)}{dt} = kS(t), \tag{4}$$

where $k > 0$.

It is a slight modification of the eqn (1). Here, the growth rate is not constant but directly proportional to the size of the growing entity. The solution of this equation, which can be found by substitution $S(t) = Z^{-1}(t)$, is given by the following simple formula:

$$S(t) = \frac{1}{a - kt}. \qquad (5)$$

It is just a reciprocal of a linearly-decreasing function. Consequently,

$$\frac{1}{S(t)} = a - kt \qquad (6)$$

The reciprocal values of the size of the growing entity follow a decreasing straight line. This representation simplifies the analysis of hyperbolic distributions. We can use this dependence to identify uniquely hyperbolic growth, in much the same way as the linearly increasing logarithm of the growing entity can be used to identify exponential growth.

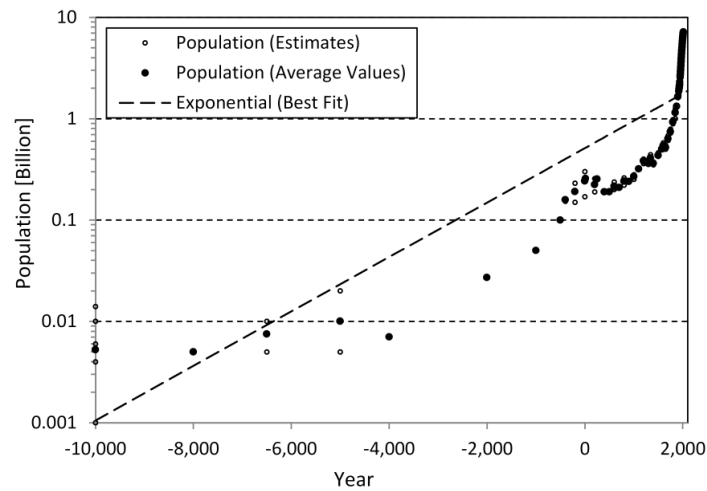

**Figure 1.** Data describing the growth of the world population (Manning, 2008; US Census Bureau, 2016) are compared with the best fit using exponential function. The world population was not increasing exponentially. The BC time scale is identified by the negative numbers.

It is now useful to understand the difference between the exponential growth and the hyperbolic growth. For the exponential growth, the growth rate is constant. It does not matter how large is the size of the growing entity, the growth rate never changes. For this reason, exponential growth can be characterised and identified by using the growth rate or equivalently by using the doubling time. This approach is inapplicable to the hyperbolic growth or to any other type of growth, for that matter. That is why it is incorrect to use the doubling time to characterise any other type of growth. In particular, it is incorrect to use the so-called "rule of 70" for any other type of growth because in all other cases the growth rate and the doubling time are not constant. In order to characterise any other types of growth by the growth rate or by the doubling time we cannot just present a single value for any of this two quantities at a certain time but we have to show how their growth rate or the doubling time depends on time or on the size of the growing entity. For instance if we look at the eqn (4) we can see that, for the hyperbolic growth, the growth rate is *directly* proportional to the size of the growing entity. This is a useful characteristic feature of hyperbolic growth.

Another characteristic feature of hyperbolic growth is that the growth rate *per size* of the growing entity is constant.

As discussed elsewhere (Nielsen, 2014), analysis and interpretation of hyperbolic distributions is difficult because they appear to be made of two distinctly-different components, slow and fast, leading to countless misconceptions and misinterpretations of hyperbolic distributions describing the growth of human population or the economic growth. However, the analysis of these distributions and their interpretation becomes trivially simple if the reciprocal values are used, as shown in Figure 2, because according to the eqns 5 and 6, if data follow a decreasing straight line, then the growth is hyperbolic. We can then fit the reciprocal values to find the mathematical expression for the hyperbolic growth given by the eqn (5).

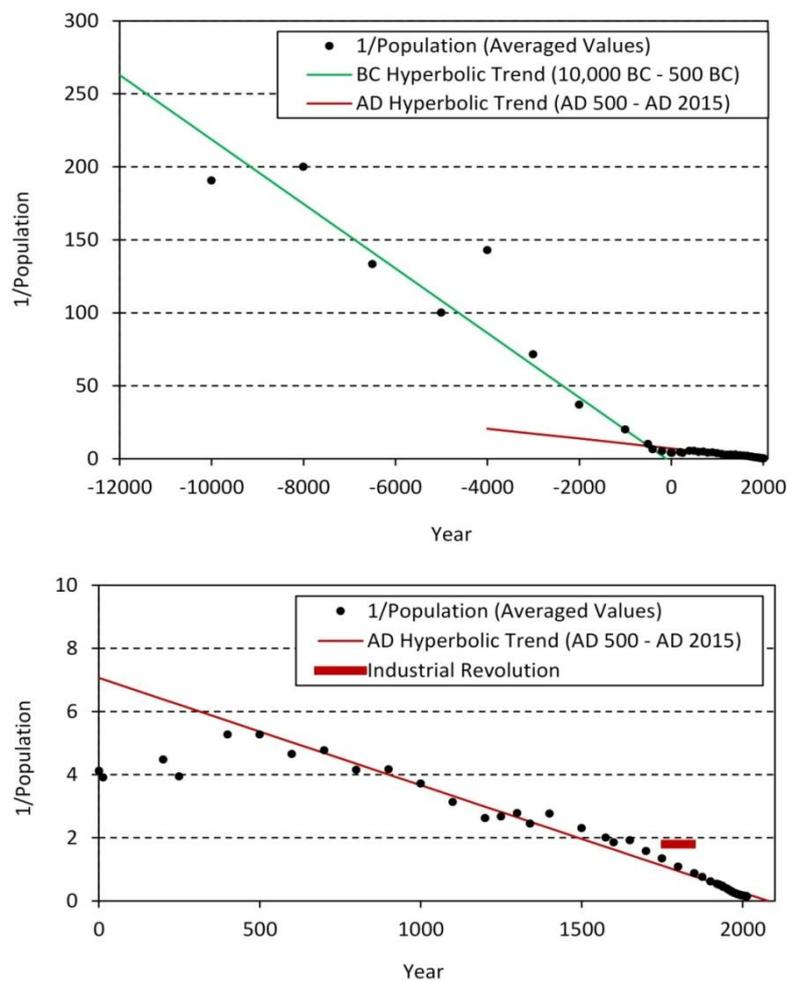

**Figure 2**. Reciprocal values of the world population data (Manning, 2008; US Census Bureau, 2015) reveal two distinctly different hyperbolic trajectories (represented by the decreasing straight lines). They also show a dramatic demographic transition between around 500 BC and AD 500. Furthermore, they show that there was no takeoff around the time of the Industrial Revolution. In fact, there was no transition from stagnation to growth at any time. The size of the population is in billions.

Furthermore, if the reciprocal values of data follow a decreasing straight line, the growth is not stagnant but hyperbolic. However, the concept of stagnation is not supported even if the reciprocal values of data do not decrease linearly. Any monotonically-decreasing trajectory will show that the postulate of stagnation followed by a takeoff at the certain time is not supported by data. To prove the existence of the epoch of stagnation it is necessary to prove the presence of random fluctuations often described as Malthusian oscillations. Such random fluctuations should be clearly seen not only in the direct display of data but also in the display of their reciprocal values. It they are absent then there is no support in data for claiming the existence of the epoch of stagnation. However, if the reciprocal values of data follow a decreasing straight line, then they show, or at least strongly suggest, that the growth was hyperbolic. Positive identification of any type of growth depends on the range of available data.

It should be also remembered that for the reciprocal values, the effects are reversed. A diversion to a *slower* trajectory will be indicated by an *upward* bending away from the earlier trajectory, while diversion to a *faster* trajectory will be indicated by the *downward* bending. Descriptions of the economic growth involve frequent discussions of the so-called takeoffs (Galor, 2005, 2011) representing the assumed sudden and prominent change in the growth trajectory, a transition from the alleged stagnation to growth. For the economic growth or for the growth of human population represented by their reciprocal values, such sudden takeoff should be indicated by a clear and strong downward bending of the growth trajectory.

If the straight line representing the reciprocal values of data remains unchanged, then obviously there is no change in the mechanism of growth. It makes no sense to divide a straight line into two or three arbitrarily selected sections and claim different regimes of growth controlled by different mechanisms for these arbitrarily-selected sections.

The analysis of data presented in Figure 2 reveals two distinctly different hyperbolic trajectories for the BC and AD eras. They are represented by two distinctly different straight lines fitting the reciprocal values of population data. In this representation, the growth during the AD era is dwarfed by the growth during the BC era but this part can be better examined by looking at the lower section of Figure 2.

The corresponding hyperbolic distributions are shown in Figure 3. Figures 1 and 2 make it clear that the growth of human population was not exponential, as it was expected by Malthus (1798). The data and their analysis show that *if unchecked, population increases hyperbolically*. It shows that the growth of human population was increasing hyperbolically not only during the AD era, as observed by von Foerster, Mora and Amiot (1960), but also during the BC era. This analysis shows also that the Industrial Revolution, 1760-1840 (Floud & McCloskey, 1994) did not boost the growth of human population, the result being in agreement with the analysis of the historical economic growth (Nielsen, 2016).

Results presented in Figures 2 and 3 show that from 10,000 BC to around 500 BC the growth of human population was hyperbolic. This hyperbolic growth was followed by a demographic transition between 500 BC and AD 500 from a fast BC hyperbolic trajectory to a significantly slower AD hyperbolic trajectory. It was not a transition from stagnation to growth because there was no stagnation in the growth of human population (Nielsen, 2013a).

Hyperbolic parameters fitting the world population data are: $a = -2.282$ and $k = 2.210 \times 10^{-2}$ for the BC trajectory between 10,000 BC and 500 BC, and $a = 7.061$ and $k = 3.398 \times 10^{-3}$ for the AD trajectory between AD 500 and 2015. Characterised by the parameter $k$, the BC hyperbolic growth was 6.5 times faster than the AD growth.

Using the data (Manning, 2008; US Census Bureau, 2016), the fitted hyperbolic distributions (shown in Figure 3) and the eqn (4) we can now estimate the growth rate during the BC and AD eras. During the BC era, the growth rate was increasing hyperbolically (monotonically) with time or linearly (and again monotonically) with the size of the population from around $1.010 \times 10^{-4}$ (0.010%) per year in 10,000 BC to around $2.520 \times 10^{-3}$ (0.252%) per year in 500 BC. The growth was slow but not stagnant. During the AD era, the growth was again approximately hyperbolic from AD 500 to 1950, and the growth rate increased approximately monotonically from $6.337 \times 10^{-4}$ (0.063%, smaller than in 500 BC) and $7.805 \times 10^{-3}$ (0.781%) in 1950.

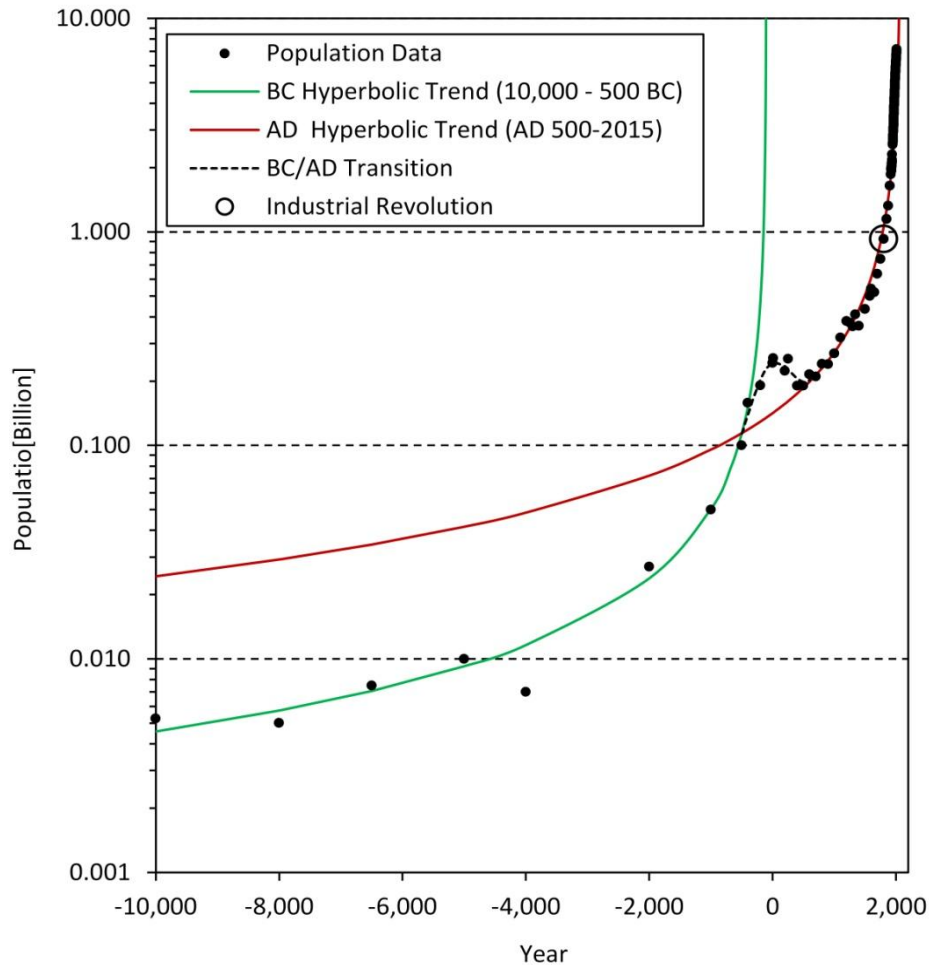

**Figure 3.** If unchecked, population increases hyperbolically. This overall view shows that there was only one major demographic transition (between around 500 BC and AD 500) from a fast to a significantly slower hyperbolic trajectory. Industrial Revolution had no impact on the growth of human population. The perceived population explosion is the natural continuation of hyperbolic growth.

There was also no stagnation but a hyperbolic growth. The transition between 500 BC and AD 500 was not a transition from stagnation to growth but from growth to growth. It was not a dramatic takeoff but a transition to a slower hyperbolic trajectory. These features are important in relating the growth of the population to the economic growth because contrary to the repeated claim in the Unified Growth Theory (Galor, 2005, 2010) there was also no dramatic takeoff in the growth of the GDP (Nielsen, 2016) or in the growth of the GDP/cap (Nielsen, 2015a).

**Detailed analysis of the AD data**

The data for the AD era are of exceptionally good quality and they allow for a closer and minute examination of the pattern of growth. Even though the hyperbolic trajectory shown in Figures 2 and 3 fits the AD data well, the display of the reciprocal values presented in the lower part of Figure 2 shows that starting from around AD 1400, some data are systematically above the fitted straight line, suggesting a shift in the hyperbolic growth around that time.

Reciprocal values of data shown in Figure 4 reveal a clear delay in the growth of the population between around AD 1200 and 1400 followed by a new and slightly faster hyperbolic trajectory. Hyperbolic trajectory between AD 500 and 1200 is given by $a = 6.940$ and $k = 3.448 \times 10^{-3}$, and from AD 1400 by $a = 9.123$ and $k = 4.478 \times 10^{-3}$. For these new and improved fits to the data, growth rate was $6.610 \times 10^{-4}$ (0.066%) in AD 500, $1.230 \times 10^{-3}$ (0.123%) in AD 1200, $1.568 \times 10^{-3}$ (0.157%) in AD 1400 and $1.142 \times 10^{-2}$ (1.142%) in 1950. The growth was hyperbolic (monotonic) between AD 500 1200 and again between AD 1400 and 1950. There was no stagnation and no dramatic takeoff from stagnation to growth at any time.

Transition between AD 1200 and 1400 coincides with the unusual convergence of strong and lethal events, representing a *combined impact of five* significant demographic catastrophes (Nielsen, 2013b): Mongolian Conquest (1260-1295) with the total estimated death toll of 40 million; Great European Famine (1315-1318), 7.5 million; the 15-year Famine in China (1333-1348), 9 million; Black Death (1343-1352), 25 million; and the Fall of Yuan Dynasty (1351-1369), 7.5 million. This is the only evidence in the data that demographic catastrophes might have had influence on the growth of the world population and if such is the case, not one but five of them were need to generate a small distortion.

There is no indication that exogenous conditions after AD 1400 were different than before AD 1200 so the slightly faster hyperbolic growth from around AD 1400 could be explained by the natural human response to crisis manifested in the intensified process of regeneration (Malthus, 1798; Nielsen, 2013c).

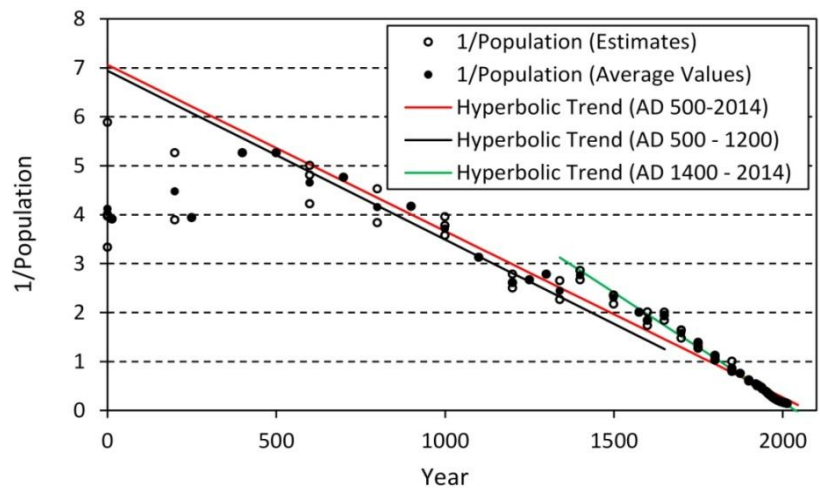

**Figure 4**. Reciprocal values of data for the AD era show a clear but small disturbance in the growth of the population between AD 1200 and 1400. This disturbance caused a shift to a slightly faster hyperbolic trajectory. The size of the population is in billions.

Closer view of the new growth trajectory, starting from around AD 1400, is displayed in Figure 5. The new hyperbolic growth was undisturbed until around 1950 when it experienced a *small but unsustained* acceleration, as indicated by a slight downward bending of the trajectory of the reciprocal values. This minor boosting lasted for only a short time and soon the growth of human population started to be diverted to a slower trajectory, as indicated by the conversion of the temporary downward bending to upward bending of the trajectory of reciprocal values.

Again, there was no dramatic takeoff and no transition from stagnation to growth, the term used repeatedly by Galor (2005, 2011) and the feature, which was supposed to characterise not only the economic growth but also the growth of human population. The repeated claim of a dramatic transition (takeoff) from stagnation to growth is contradicted by the analysis of the economic growth (Nielsen, 2015a, 2016) and by the presented here analysis of the growth of the world population.

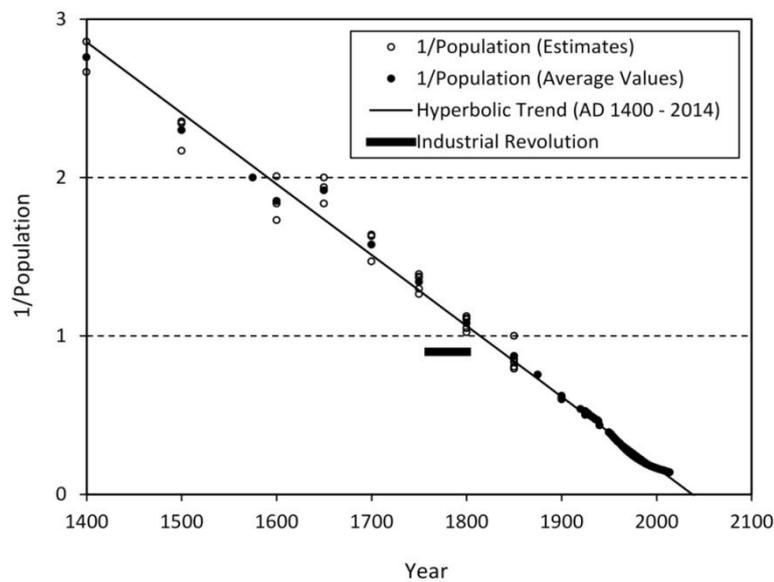

**Figure 5**. Between AD 1400 and around 1950 the growth of human population was hyperbolic. Data show a minor boosting around 1950 followed quickly by a diversion to a slower trajectory. There was no takeoff from stagnation to growth at any time. Industrial Revolution had no impact on boosting the world population. The size of the population is in billions.

It is remarkable that the growth of the world population was so hyperbolically stable over the past 12,000 years. The data show that during this long time that there were only three transitions: 500 BC - AD 500, AD 1200 - AD 1400 and 1950 - present. Each of the two earlier transitions was a shift between hyperbolic trajectories. The outcome of the current transition is unknown. The dynamics of growth in the past 12,000 years is summarised in Table 1.

**Table 1.** Dynamics of growth of the world population in the past 12,000 years. Time intervals are approximate.

| Hyperbolic Growth | Demographic Transitions |
|---|---|
| 10,000 BC – 500 BC | 500 BC – AD 500 |
| $a = -2.282$; $k = 2.210 \times 10^{-2}$ | Transition from a fast to much slower hyperbolic trajectory |
| AD 500 – 1200 | AD 1200 – 1400 |
| $a = 6.940$; $k = 3.448 \times 10^{-3}$ | Transition from a slow to a slightly faster hyperbolic trajectory |
| AD 1400 – 1950 | 1950 – present |
| $a = 9.123$; $k = 4.478 \times 10^{-3}$ | Transition from a hyperbolic trajectory to an unknown trend |
| Total time of hyperbolic growth: 10,750 years (~89% of the total combined time) | Total rime of transitions 1265 years (~11% of the total combined time) |

**Implications for the economic growth**

As mentioned earlier, preliminary analysis of Maddison's data (Maddison, 2001, 2010) shows close similarities between the distributions describing economic growth and the growth of human population. Galor also commented that there was a "positive relationship between income per capita and population that existed throughout most of human history" (Galor, 2005, p. 177). The study of the economic growth goes hand in hand with the study of the growth of the population.

Our analysis demonstrated that the growth of the world population was hyperbolic, and consequently monotonic, and that there was never a transition from stagnation to growth, which could be described as a sudden takeoff. The fast increasing growth of the world population in recent years was just the natural continuation of the hyperbolic growth.

Our analysis shows that with the exception of just two demographic transitions (500 BC - AD 500, and AD 1200 - 1400) the growth of human population was monotonic until around 1950, when it started to be diverted to a yet unknown trajectory. The first demographic transition (500 BC - AD 500) was from a faster to a slower hyperbolic growth. It was definitely not a takeoff from stagnation to growth. The second transition (AD 1200 – 1400) was from a slow to a slightly faster hyperbolic trajectory (only 30% faster, as indicated by the parameter *k*). It was also not a transition from stagnation to growth. The current transition, which commenced around 1950 was initially to a slightly faster trajectory, which was soon becoming progressively slower than the preceding hyperbolic trajectory. Here again, there was no transition from stagnation to growth. For 89% of the past 12,000 years the growth of human population was hyperbolic and monotonic and there was never a transition from stagnation to growth. Our analysis shows that the growth of human population was remarkably stable over the past 12,000 years.

Galor wonders "what is the origin of the sudden spurt in growth rates of output per capita and population?" (Galor, 2005, p. 177). This puzzle has now been solved: *there was no sudden spurt*.

Trying to explain this sudden spurt is like trying to explain why there is water in the middle of the desert when the image of water is created by a mirage. It is a waste of time and effort. We can explain the *illusion* of the spurt but not the spurt. The illusion of the spurt is explained by the hyperbolic properties but the sudden spurt has never happened. What we see as a sudden spurt is the natural continuation of the monotonically-increasing hyperbolic distribution and the simplest way to dispel the illusion of stagnation and of a sudden spurt is to use the reciprocal values of data (Nielsen, 2014) but we can also use other methods (Nielsen, 2015a). The point is that data have to be rigorously analysed. Any perfunctory and hasty examination of data is likely to lead to incorrect conclusions and we can find many examples of such examinations of data in the Unified Growth Theory (Galor, 2005, 2011).

We have demonstrated that there was no sudden spurt in the growth rate of the world population because the growth was hyperbolic, which means that the growth rate was also increasing hyperbolically with time or linearly with the size of the population, in both cases monotonically [see the eqn (4)]. Such an increase has no room for any form of spurts.

There were also no spurts during the past two demographic transitions. During the first transition (500 BC - AD 500), the growth rate decreased from 0.252% in 500 BC to 0.066% in AD 500. During the second transition (AD 1200 - 1400) the growth rate increased only slightly from 0.123% in AD 1200 to 0.157% in AD 1400.

So, our analysis eliminates at least one of Galor's spurts: the alleged spurt in the growth rate of human population. What remains to be explained is the alleged spurt in the growth rate of output per capita (GDP/cap) but the analysis of this ratio shows that the growth rate of the GDP/cap was also increasing monotonically (Nielsen, 2015a). There was no spurt at all. Furthermore, the analysis of the GDP data (Nielsen, 2016) also shows that there were no spurts (takeoffs) in the growth of the GDP.

When data are closely analysed they show that what Galor saw as spurts in the growth rates represented just the natural features of monotonically increasing hyperbolic distributions describing the growth of the population, the growth of the GDP and of the growth of the GDP/cap, and of their respective monotonically-increasing growth rates. All these distributions were slow over a long time and fast over a short time. These features are real but they represent nothing mysterious but the natural properties of monotonically-increasing hyperbolic distributions. They create strong illusions of stagnations followed by sudden spurts or takeoffs but when properly analysed they show that there was no stagnation and that the sudden spurts (takeoffs) never happened.

Galor wonders about the relationship between the income per capita (GDP/cap) and the population growth, but the answer to this apparent riddle is simple. When closely analysed, the growth of the population is found to be hyperbolic. The growth of the GDP is also hyperbolic (Nielsen, 2016) and hence, the growth of the GDP/cap is described by the ratio of hyperbolic distributions, which is just a linearly-modulated hyperbolic distribution (Nielsen, 2015a). The mystery is solved.

The only features, which need to be explained, are not the stagnation and sudden spurts (takeoffs) because they did not exist but why the growth of human population and the growth of the GDP were hyperbolic. This issue diverts our attention from phantom problems, which do not need to be solved, and directs it to the problem, which needs to be solved, because if we could explain why the growth of the population and the growth of the GDP were hyperbolic, we could also explain the time dependence of the historical income per capita.

Finally, we shall address a minor issue, which might help to understand at least one discrepancy between the fitted hyperbolic curve and the GDP data (Nielsen, 2016). In that analysis we have found that one point, located at AD 1 was 77% higher than the fitted hyperbolic distribution. In Figure 3 we can see that something similar can be observed for the growth of human population. The size of the population in AD 1 was 71% higher than the size determined by the fitted hyperbolic distribution to the AD data, and the explanation of this discrepancy is simple: there was a maximum in the growth of the population around AD 1 caused by the transition from a fast hyperbolic trajectory during the BC era to a significantly slower hyperbolic trajectory during the AD era. Close similarities between the growth of the GDP and the growth of the population displayed by Maddison's data (Maddison, 2001, 2010) suggest that the 77% difference between the GDP value and the fitted hyperbolic distribution at AD 1 (Nielsen, 2016) might reflect a similar maximum in the growth of the GDP as observed in the growth of the population.

**Summary and conclusions**

We have analysed the world population data (Manning, 2008; US Census Bureau, 2015) between 10,000 BC and AD 2015. We have found that the growth was hyperbolic during the BC and AD eras.

We have also found that there were just three, relatively, brief demographic transitions during that time: between 500 BC and AD 500, between AD 1200 and 1400 and currently from around 1950. These transitions were of a different kind than usually discussed in academic publications. None of them was a transition from stagnation to a fast growth. None of them represented a sudden takeoff from stagnation to growth, the feature discussed extensively in the Unified Growth Theory (Galor, 2005, 2011).

The first transition was from a fast hyperbolic trajectory to a significantly slower hyperbolic trajectory; the second from a slow hyperbolic trajectory to a slightly faster hyperbolic trajectory; and the current transition from the latest hyperbolic trajectory to a yet unknown trend. The total fraction of time characterising hyperbolic growth was about 89% of the past 12,000 years and the total time taken by transitions was only about 11%. Thus the analysis shows that if unchecked, population does not increase exponentially as believed by Malthus but hyperbolically. There was also no stagnation in the growth of the world population (Nielsen, 2013a), not only during the AD era but also during the BC era.

Correct understanding of the growth of human population is essential for the correct understanding of economic growth because, as pointed out by Galor (2005, 2011) there is a close relationship between the growth of the population and the growth of income per capita (GDP/cap). We have demonstrated that the growth of the world population was hyperbolic. The growth of the world GDP/cap can be also described using hyperbolic distributions. It is simply a ratio of the hyperbolic distribution describing the growth of the world GDP and the hyperbolic distribution describing the growth of human population (Nielsen, 2015a). Furthermore, it has been already shown that the regional growth of the GDP was hyperbolic (Nielsen, 2016). Similar study could be extended to the growth of regional populations. However, what is already becoming clear is that in order to explain the mechanism of the historical economic growth, expressed either as the GDP or GDP/cap, our attention should be diverted from trying to explain phantom features of stagnation and takeoffs, discussed so extensively in the Unified Growth Theory (Galor, 2005, 2011), the features that did not exist, and that our efforts should be focused on explaining why the economic growth and the growth of human population were hyperbolic.

**References**

Durand, J. D. (1977). Historical estimates of world population: An evaluation. *Population and Development Review, 3*(3), 253-296.

Floud, D. & McCloskey, D.N. (1994). *The Economic History of Britain since 1700*. Cambridge: Cambridge University Press.

Galor, O. (2005). From stagnation to growth: Unified Growth Theory. In P. Aghion & S. Durlauf (Eds.), *Handbook of Economic Growth* (pp. 171-293). Amsterdam: Elsevier.

Galor, O. (2011). *Unified Growth Theory*. Princeton, New Jersey: Princeton University Press.

Maddison, A. (2001). *The World Economy: A Millennial Perspective*. Paris: OECD.

Maddison, A. (2010). Historical Statistics of the World Economy: 1-2008 AD. http://www.ggdc.net/maddison/Historical Statistics/horizontal-file_02-2010.xls.

Malthus, T. R. (1798). *An Essay on the Principle of Population as It Affects the Future Improvement of Society, with Remarks on the Speculations of Mr Godwin, M. Condorcet, and Other Writers*. London: J. Johnson.

Manning, S. (2008). *Year-by-Year World Population Estimates: 10,000 B.C. to 2007 A.D*. http://www.scottmanning.com/content/year-by-year-world-population-estimates/ and references therein.

Nielsen, R. W. (2013a). No stagnation in the growth of population. http://arxiv.org/ftp/arxiv/papers/1311/1311.3997.pdf

Nielsen, R. W. (2013b). *Impact of demographic catastrophes*. http://arxiv.org/ftp/arxiv/papers/1311/1311.1850.pdf

Nielsen, R. W. (2013c). *Malthusian stagnation or Malthusian regeneration?* http://arxiv.org/ftp/arxiv/papers/1310/1310.5390.pdf

Nielsen, R. W. (2014). Changing the Paradigm. *Applied Mathematics*, *5*, 1950-1963. http://dx.doi.org/10.4236/am.2014.513188

Nielsen, R. W. (2015a). Unified Growth Theory Contradicted by the GDP/cap Data. http://arxiv.org/ftp/arxiv/papers/1511/1511.09323.pdf

Nielsen, R. W. (2016). Mathematical analysis of the historical economic growth with a search for takeoffs from stagnation to growth, *Journal of Economics Library*, 3(1), *forthcoming*.

US Census Bureau (2016). *International Data Base.* http://www.census.gov/ipc/www/idb/worldpopinfo.php and references therein.

von Foerster, H., Mora, P., & Amiot, L. (1960). Doomsday: Friday, 13 November, A.D. 2026. *Science, 132*, 255-296.